\documentclass{elsart}

\usepackage{graphicx}
\usepackage{subfigure}

\usepackage{amssymb}

\begin{document}

\begin{frontmatter}


\title{Two Mutually Loss-coupled Lasers Featuring Astable Multivibrator}
\author{A. Mustafin}
\address{Department of Physics, University of Maryland, College Park, MD 20742, USA}





\begin{abstract}
Self-sustained antiphase relaxation oscillations of high amplitude
are shown to be possible in a system of two single-mode
semiconductor lasers strongly coupled through their cavities.
\end{abstract}

\end{frontmatter}

\section{Introduction}
\label{intro} It has been known that individual single mode laser
does not exhibit self-sustained oscillations. However self-pulsing
regimes of different complexity are well possible in coupled lasers.
Recently synchronization in coupled semiconductor lasers has been a
subject of extensive study not only because laser is one of the
important nonlinear dynamical systems, but also owing to potential
applications of synchronization phenomena to communication,
electronic circuits and even biological systems.

The overwhelming majority of synchronization schemes employ either
various modifications of optoelectronic cross-coupling
\cite{optoelectronic}, or face-to-face mutual coupling
\cite{face-to-face}. In the former scheme, the output of each laser
is detected and converted into an electronic signal by a
photodetector, and after amplification the signal is fed back to
modulate the pump current of the other laser. In the latter scheme,
the output of each laser is injected, after a suitable attenuation,
in the other laser. While handling the above two types of coupling
both theoretically and experimentally in numerous papers, a whole
series of interesting effects has been discovered. Under certain
conditions, a time delay introduced by the mutual feedback being
paramount, coupling can drive semiconductor lasers into nonlinear
oscillations, such as regular pulsing, quasi-periodic pulsing, or
chaotic pulsing.

At the same time, there is perhaps the only work dealing with an
alternative mechanism of intensity coupling based on cross-loss
\cite{Nguyen1999}. In their version of such a coupling scheme, an
additional loss in either laser is induced by macroscale mechanical
deformations of the crystal structure in a common support due to
strong local heating. Consideration based on the equations proposed
by \cite{Tang1963} with added intensity-dependent losses has shown a
self-pulsing instability of the steady state solutions.

There is good reason to believe that cross-loss coupling harbors a
great deal of interesting synchronization effects highly competitive
in diversity with those of the known coupling schemes. However the
ther\-mo\-me\-cha\-ni\-cal embodiment of this principle suffers from
the slowness of heating in relation to the dynamics of photons and
carriers. Signal exchange via ther\-mo\-me\-cha\-ni\-cal modulation
of the crystal structure cannot be sped up by placing the beams
closer to each other because experimentalist has to avoid
field-field interaction. More promising way is to use one or other
type of intra\-ca\-vi\-ty q-modulator controlled by electric,
magnetic, or acoustic pulses.

In the present paper a system of two semiconductor lasers coupled in
such a manner that cavity loss rate of each one is proportional to
the output of its counterpart, is studied in terms of the rate
equations for photon and carrier densities. Another key assumption
of the model is existence of a nonzero second-order cavity loss
associated with two-photon absorption at high powers. The plausible
range for that second-order loss rate is estimated. The existence of
a self-pulsing regime featured by spiky antiphase relaxation
oscillations is inferred from the system of coupled rate equations
by applying multiple-scale approximation techniques. Any time delay
in coupling is not a prerequisite for the emergence of synchronous
oscillations. The frequency of the obtained antiphase-locked pulsing
is shown to be considerably lower than that of intrinsic underdamped
quasi-harmonic oscillations of a standalone laser and completely
independent of the concrete value of the second-order loss rate.

\section{Rate equations of a single-mode laser}
\label{standalone} We take, as the starting point, the following
rate equations for a single-mode semiconductor laser:
\begin{equation} \label{rate eqs}
\begin{array}{rcl} \dot{P} & = & \left( G(N)-\gamma_p-D P \right) P, \\
\dot{N} & = & \displaystyle \frac{J}{e d}-G(N) P-\gamma_n N.
\end{array}
\end{equation}
(The dot denotes $d/dt$.) Here $P$ and $N$ are the respective
densities of photon and carrier population inversion inside the
laser cavity. The cavity is assumed to have thickness $d$. Linear
function of carriers, $G(N)=\Gamma v_{\mathrm{g}} a (N-N_0)$, is the
net rate of stimulated emission, where $\Gamma$ is the confinement
factor (the ratio of the volume of the cavity to the volume occupied
by photons in the cavity), $v_{\mathrm{g}}$ is the light group
velocity, $a$ is the gain constant, and $N_0$ is the carrier density
at transparency. $\gamma_p=v_{\mathrm{g}} (\alpha_{\mathrm{f}} +
\alpha_{\mathrm{int}})$ is the photon loss rate due to both facet
($\alpha_{\mathrm{f}}$) and internal ($\alpha_{\mathrm{int}}$)
losses. $J$ is the density of a pump current flowing through the
active region and $e$ is the elementary charge. $\gamma_n$ is the
carrier loss rate solely due to nonradiative effects. Somewhat
simplifying the picture, we assume the contribution of both
radiative and Auger recombination loss mechanisms negligible. Taking
those into account does not qualitatively affect our ensuing
results, however makes consideration more involved.

An additional (small) quadratic loss term $DP^2$ is introduced in
the first equation of (\ref{rate eqs}) to allow for nonlinear
mechanism of photon fluctuations damping, such as two-photon
absorption, anticipated at high powers.

The meaning and typical values of the different parameters (mostly
borrowed from \cite{Agrawal-Dutta1993}) in the model (\ref{rate
eqs}) are given in Table \ref{tab1}. The same numerical values of
the parameters are used in the calculations. {\small \begin{table}
\caption{Typical parameter values for a single-mode uncoupled laser}
\label{tab1}
\begin{tabular}{llll} \hline
Parameter & Meaning & Value & Units \\
\hline
$c$ & speed of light in vacuum & $3{\cdot}10^{10}$ & cm/s \\
$e$ & elementary charge & $1\ldotp 6{\cdot}10^{-19}$ & C \\
$\mu_{\mathrm{g}}$ & group refractive index & 4 & --- \\
$v_{\mathrm{g}}=c/\mu_{\mathrm{g}}$ & light group velocity & $0\ldotp 75{\cdot}10^{10}$ & cm/s \\
$\Gamma$ & confinement factor & $0\ldotp 3$ & --- \\
$a$ & gain constant & $2\ldotp 5{\cdot}10^{-16}$ & $\mathrm{cm}^2$ \\
$d$ & cavity thickness & $2{\cdot10}^{-5}$ & cm \\
$N_0$ & carrier density at transparency & $10^{18}$ & $\mathrm{cm}^{-3}$ \\
$\gamma_n$ & nonradiative carrier loss rate & $10^8$ &
$\mathrm{s}^{-1}$ \\
$\alpha_{\mathrm{f}}$ & facet cavity loss & 45 & $\mathrm{cm}^{-1}$ \\
$\alpha_{\mathrm{int}}$ & internal cavity loss & 40 & $\mathrm{cm}^{-1}$ \\
$\gamma_p$ & cavity loss rate & $0\ldotp 6375{\cdot}10^{12}$ & $\mathrm{s}^{-1}$ \\
$J$ & pump current density & $5{\cdot}10^{3}$ & $\mathrm{A}/\mathrm{cm}^2$ \\
$j$ (eq. (\ref{scaling})) & dimensionless pump & $11\ldotp 90$ & --- \\
$D$ & second-order cavity loss & $2\ldotp 954{\cdot}10^{-6}$ &
$\mathrm{cm}^3/\mathrm{s}$ \\
& & (conditional) & \\
$\delta$  (eq. (\ref{scaling}))&
dimensionless second-order cavity loss & $0\ldotp 8239{\cdot}10^{-3}$ & --- \\
$\varepsilon$ (eq. (\ref{scaling}))& ratio of the time constants &
$1\ldotp 569{\cdot}10^{-4}$ & --- \\
$T_{\mathrm{intr}}$ (eq. (\ref{intr freq})) & period of intrinsic
oscillations &
$2\ldotp 281{\cdot}10^{-10}$ & $\mathrm{s}$ \\
\hline
\end{tabular}
\end{table}}
Equations (\ref{rate eqs}) can be converted into dimensionless form
by performing the linear scaling
\begin{equation} \label{scaling}
\begin{array}{rclrcl}
p & = & P \Gamma v_{g} a \gamma{}_{n}^{-1}, \qquad & n & = &
(N-N_{0})
\Gamma v_{g} a \gamma{}_{p}^{-1} -1,\\
\delta & = & D \gamma_{n}(\Gamma v_{g} a \gamma_{p})^{-1}, \qquad &
j & = & \left( J (e d \gamma{}_{n})^{-1} -N_{0} \right) \Gamma v_{g}
a \gamma{}_{p}^{-1} - 1,\\
\varepsilon & = & \gamma_n/\gamma_p\,,&&&\\
\end{array}
\end{equation}
and they become
\begin{equation} \label{basic eqs}
\begin{array}{rcl} \dot{p} & = & \varepsilon^{-1} \gamma_n (n-\delta p)p, \\
\dot{n} & = & \gamma_n \left( j-(n+1)p-n \right).
\end{array}
\end{equation}
Note that now the dimensionless $n$ and $j$ are no longer
proportional to their dimensional prototypes, but rather are
deviations of the corresponding absolute quantities from the
threshold of generation. Value of $\varepsilon$ representing the
ratio of the time constants of the two equations is of order
$O(10^{-4})$ because the rate constants $\gamma_p$ and $\gamma_n$
differ almost ten thousand times.

Seemingly transparent, the dynamics of system (\ref{basic eqs}) is
worth touching briefly on to estimate how small the hypothetical
second-order cavity loss might be and to provide more seamless
passage to the coupled dynamics to be treated in the next section.

Two physically meaningful steady states of system (\ref{basic eqs})
are possible in the $(p,\,n)$ phase plane (Fig. \ref{single laser}):
\begin{equation}
\label{solutions}
\begin{array}{rcl}
p^{(1)} & = & 0, \qquad  n^{(1)} = j;\\
p^{(2)} & = & \left( \sqrt{(1+\delta)^2+4 \delta j}\,-1-\delta
\right)/(2 \delta) = j-\delta j (j+1) + O(\delta^2),\\
n^{(2)} & = & \left( \sqrt{(1+\delta)^2 + 4 \delta j}\,-1-\delta
\right)/2 = \delta j + O(\delta^2).
\end{array}
\end{equation}
\newcommand{\tr}{\mathop{\mathrm{tr}}}
Steady state 1 is always a saddle since eigenvalues $\lambda$ of the
Jacobian matrix $\mathbf{J}$ of (\ref{basic eqs}), being the roots
of the characteristic polynomial $\det{(\mathbf{J}-\lambda
\mathbf{I})}=\lambda^2 - (\tr{\mathbf{J}}) \lambda +
\det{\mathbf{J}}$, have opposite signs: $\lambda_{1}=-\gamma_{n}$,
$\lambda_{2}=j \gamma_{n}/\varepsilon$.

Steady state 2 is always a stable focus/node, because
$\tr{\mathbf{J}} = -1 -p^{(2)} -\varepsilon^{-1} \delta p^{(2)} < 0$
and $\det{\mathbf{J}} = \varepsilon^{-1} \gamma{}_{n}^2 p^{(2)}
\left( 1 + \delta (1 + 2 p^{(2)} ) \right) > 0$.
\begin{figure}
\begin{center}
\subfigure[]{\includegraphics[width=0.48\textwidth]{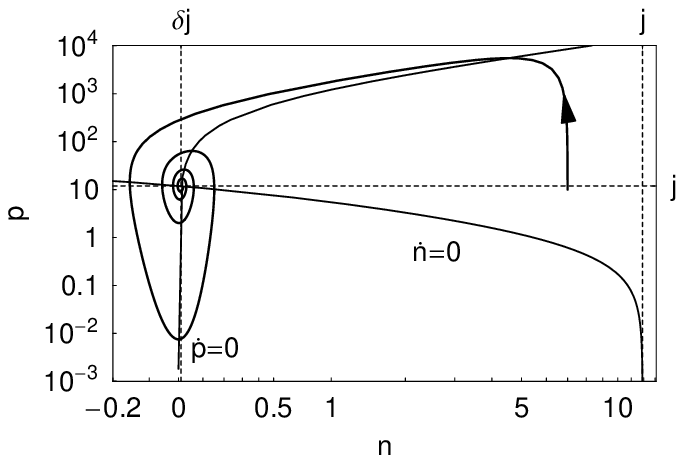}}\hfill
\subfigure[]{\includegraphics[width=0.48\textwidth]{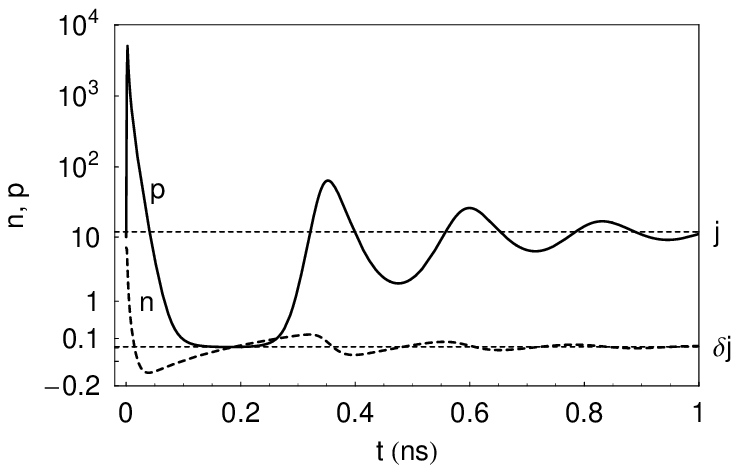}}
\caption{(a) Phase portrait of a single laser corresponding to
system (\ref{basic eqs}). Steady states lie at the intersections of
nullclines $\dot{p}=0$ and $\dot{n}=0$. Steady state with nonzero
population inversion is a stable focus, while the steady state on
the $n$-axis is a saddle. (b) Time profiles of underdamped intrinsic
oscillations (solid: photons, dashed: carriers).} \label{single
laser}
\end{center}
\end{figure}
Normally, it must be a focal point, because it is just transient
intrinsic underdamped oscillations that could be experimentally
observed in a semiconductor laser briefly after turning it on
\cite{relax}. For steady state 2 to be a focus, the discriminant of
the quadratic characteristic equation has to be negative:
$\Delta=(\tr{\mathbf{J}})^2-4\det{\mathbf{J}}<0$. Correct to the
second order in $\delta$, this condition can be reduced to
\begin{displaymath}
\Delta=\varepsilon^{-2} \gamma_{n}^{2} j (\delta^2
j-4\varepsilon)<0.
\end{displaymath}
Solving the above inequality yields an upper bound for $\delta$:
\begin{equation}
\label{upper bound} \delta < \delta_\mathrm{max} = 2
\sqrt{\varepsilon/j}\,.
\end{equation}
For a laser with typical $j$ and $\varepsilon$ given in Table
\ref{tab1}, the magnitude of $\delta_\mathrm{max}$ is $0\ldotp
726{\cdot}10^{-2}$.

While $\delta$ remains such that $\delta{=}o(\varepsilon^{1{/}2})$,
the second-order loss has a negligible effect on the period of the
intrinsic oscillations, for which we get
\begin{equation}
\label{intr freq} T_\mathrm{intr}=4\pi |\Delta|^{-1/2}=2 \pi
\gamma{}_{n}^{-1} \sqrt{\varepsilon/j}\,.
\end{equation}
Numerically, this is 228 ps.

The lower-bound estimate for the second-order loss can be performed
by studying the decay time of the intrinsic oscillations. The decay
time is the inverse of the damping factor, which in turn is a half
of the trace of the Jacobian matrix: $\tau_\mathrm{dec}=2
 |\tr{\mathbf{J}}|^{-1}$. Keeping only the first terms in the
expansion of $\tr{\mathbf{J}}$ in powers of $\delta$ yields
$\tau_\mathrm{dec}=2\varepsilon
 \gamma{}_{n}^{-1}\left( \delta j+\varepsilon (j+1) \right)^{-1}$.

Two different asymptotic cases are possible. In the first case
$\delta=o(\varepsilon)$.  This condition is stronger than
(\ref{upper bound}). We then obtain $\tau_\mathrm{dec}=2
 \gamma{}_{n}^{-1}(j+1)^{-1}$. The
result means that in the absence of the second-order cavity loss the
decay time is totally determined by the carriers' recombination rate
and must be of the order of several nanoseconds.

In the alternative case $\varepsilon=o(\delta)$. Now
$\tau_\mathrm{dec}=2\varepsilon (\gamma_{n} \delta j)^{-1}$, so the
intrinsic oscillations may decay $\delta{/}\varepsilon$ times
faster. This is a noticeable difference contributed by the
second-order damping. It should be mentioned, that C.J.~Kennedy and
J.D.~Barry \cite{Kennedy-Barry1974} were first to indicate the
importance of the second-order damping although the damping in their
laser system had a different physical nature, resulting from the
intracavity frequency doubling. It is precisely this interplay
between the two small parameters in the system, $\varepsilon$ and
$\delta$, that is of crucial importance in determining the temporal
hierarchy of a laser. If we are given some experimentally observed
maximum characteristic time of decay, $\tau_\mathrm{decmax}$, then
the lower bound for $\delta$ can be found from the condition
$\tau_\mathrm{dec}<\tau_\mathrm{decmax}$ resulting in
\begin{equation}
\label{lower bound} \delta>\delta_\mathrm{min} = \varepsilon \left(
2/(\gamma_{n} \tau_\mathrm{decmax})-j-1 \right)/j.
\end{equation}
It has been known from the experiments \cite{Agrawal-Dutta1993} that
intrinsic oscillations in semiconductor lasers do decay much faster
than the standard rate equations involving only first-order cavity
loss predict. We can ascribe this extra damping to the second-order
cavity loss. The observed $\tau_\mathrm{decmax}$ is usually shorter
than a nanosecond. Hence for the typical values of $j$ and
$\varepsilon$, $\delta_\mathrm{min}$ would be $0\ldotp
935{\cdot}10^{-4}$. For the present we cannot judge with any
confidence about the value of $\delta$, except of it must be related
to $\varepsilon$ via conditions (\ref{upper bound}) and (\ref{lower
bound}). Nevertheless, as we will demonstrate further, the exact
value of $\delta$ is not all that critical, and all the results
would remain valid providing this parameter is within the wide range
from (\ref{lower bound}) through (\ref{upper bound}). For the
purposes of model calculations, we adopted $\delta$ to be a
geometric mean of $\delta_\mathrm{min}$ and $\delta_\mathrm{max}$,
i.e. $0\ldotp 8239{\cdot}10^{-3}$.

An alternative way to introduce the effective second-order cavity
loss in a model is to assume that the material gain has a linear
dependence not only on the carrier but also on the photon density,
as is done in the work of R.~Vicente et al. \cite{Vicente2004}.

Since the small parameter multiplies the derivative $\dot{p}$,
system (\ref{basic eqs}) is \emph{singularly perturbed\/} (e.\,g.
\cite{OMalley1974}). The significant difference between carrier and
photon lifetimes brings multiscale properties into the model. It
contains relatively fast variable, $p$, and slow variable, $n$.
However, the hasty conclusion that $\gamma{}_n^{-1}$ and
$\gamma{}_p^{-1}$ are some unique characteristic times of
fluctuations of the corresponding variables would be an
oversimplification. Indeed, at any current value of $n$ the
fluctuations of $p$ have a characteristic time of order
$\varepsilon/n$. Thus, in the vicinity of the focal point the
difference in rates of change of the two variables is not as great
as when far from the steady state.

\section{Two loss-coupled lasers: statics}
\label{statics} Consider a pair of not necessarily identical, but
having comparable parameters, lasers of type (\ref{rate eqs})
cross-coupled through their resonators, so that each of them can
modulate cavity loss of the other:
\begin{displaymath}
\begin{array}{rcl}
\dot{P}_1 & = & \left( G_1 (N_1) - \gamma_{p_1}
- D_1 P_1 - K_2 P_2 \right) P_1, \\
\dot{P}_2 & = & \left( G_2 (N_2) - \gamma_{p_2}
- D_2 P_2 - K_1 P_1 \right) P_2, \\
\dot{N}_1 & = & J_1/(e d_1) - G_1 (N_1) P_1 -
\gamma_{n_1} N_1, \\
\dot{N}_2 & = & J_2/(e d_2) - G_2 (N_2) P_2 - \gamma_{n_2} N_2.
\end{array}
\end{displaymath}
Here $K_1$ and $K_2$ are positive coupling strengths. On the proper
rescaling, this system takes the following form:
\begin{equation}
\label{coupled lasers}
\begin{array}{rcl}
\dot{p}_1 & = & \varepsilon{}_1^{-1} \gamma_{n_1} (n_1 - \delta_1
p_1 - \varkappa_2 p_2) p_1, \\
\dot{p}_2 & = & \varepsilon{}_2^{-1} \gamma_{n_2} (n_2 - \delta_2
p_2 - \varkappa_1 p_1) p_2, \\
\dot{n}_1 & = & \gamma_{n_1} \left( j_1 - (n_1 + 1) p_1 - n_1
\right), \\
\dot{n}_2 & = & \gamma_{n_2} \left( j_2 - (n_2 + 1) p_2 - n_2
\right).
\end{array}
\end{equation}
The dimensionless coupling strengths $\varkappa_{1,\,2} = K_{1,\,2}
\gamma_{n_{1,\,2}} (\gamma_{p_{2,\,1}}\Gamma_{1,\,2}
v_\mathrm{g{1,\,2}} a_{1,\,2})^{-1}$ do not have to be weak; we
assume that at least,
\begin{equation}
\label{kappa gt gamma} \varkappa_{1,\,2} \gg \delta_{1,\,2}\,.
\end{equation}
Model (\ref{coupled lasers}) has four steady states. To $O(1)$ for
small $\delta_{1,\,2}$,
\begin{equation}
\label{coupled sol}
\begin{array}{rcl}
p{}_1^{(1)} & = & j_1, \qquad p{}_2^{(1)} = 0,\qquad
n{}_1^{(1)} = 0, \qquad n{}_2^{(1)} = j_2;\\
p{}_1^{(2)} & = & 0, \qquad p{}_2^{(2)} = j_2,\qquad
n{}_1^{(2)} = j_1, \qquad n{}_2^{(2)} = 0;\\
p{}_1^{(3)} & = & 0, \qquad p{}_2^{(3)} = 0, \qquad
n{}_1^{(3)} = j_1, \qquad n{}_2^{(3)} = j_2;\\
p{}_1^{(4)} & = & \left( 1 - \varkappa_1 (j_1 + \varkappa_2) +
\varkappa_2 j_2 - Q \right)/\left( 2 \varkappa_1 (\varkappa_2 - 1) \right)\\
& = & j_1 - \varkappa_2 j_2 (j_1 + 1) + O(\varkappa{}_1^2 + \varkappa{}_2^2),\\
p{}_2^{(4)} & = & \left( 1 - \varkappa_2 (j_2 + \varkappa_1) +
\varkappa_1 j_1 - Q \right)/\left( 2 \varkappa_2 (\varkappa_1 - 1) \right)\\
& = & j_2 - \varkappa_1 j_1 (j_2 + 1) + O(\varkappa{}_1^2 + \varkappa{}_2^2),\\
n{}_1^{(4)} & = & \varkappa_2 p{}_2^{(4)} = \varkappa_2 j_2 +
O(\varkappa{}_1^2 + \varkappa{}_2^2),\\
n{}_2^{(4)} & = & \varkappa_1 p{}_1^{(4)} = \varkappa_1 j_1 +
O(\varkappa{}_1^2 + \varkappa{}_2^2),\\
Q & = & \left( \left(1 + \varkappa_2 j_2 - \varkappa_1 (j_1 +
\varkappa_2) \right)^2 + 4 \varkappa_1 (\varkappa_2 - 1)(\varkappa_2
j_2 - j_1) \right)^{1/2}.
\end{array}
\end{equation}
Steady states 1 and 2 are {\textquotedblleft pure\textquotedblright}
in the sense that either of them corresponds to one device lasing
while the other being inactive. Correct to $O(1)$ in
$\varepsilon_{1,\,2}$, the eigenvalues for steady state 1 are
$\lambda_1 = -\gamma_{n_{2}}$, $\lambda_2 = -\varepsilon{}_2^{-1}
\gamma_{n_{2}} (\varkappa_1 j_1 - j_2)$,
$\lambda_{3,\,4}=-\gamma_{n_{1}}\left( (j_1 + 1)/2 \pm \mathrm{i}
\sqrt{j_1/\varepsilon_1}\, \right)$; whence it follows that this
equilibrium is stable when
\begin{equation}
\label{ss1 stability} \varkappa_1 > j_2/j_1.
\end{equation}
Similar reasoning shows that steady state 2 is stable when
\begin{equation}
\label{ss2 stability} \varkappa_2 > j_1/j_2.
\end{equation}
Steady states 3 and 4 are {\textquotedblleft
mixed\textquotedblright} in the sense that either of them
corresponds to both lasers being in the same mode of operation. In
steady state 3 both lasers are inactive (do not emit any light).
This steady state is always unstable, because two of the four
associated eigenvalues are positive: $\lambda_1 = -\gamma_{n_{1}}$,
$\lambda_2 = -\gamma_{n_{2}}$, $\lambda_3 = \varepsilon{}_1^{-1}
\gamma_{n_{1}} j_1$, $\lambda_4 = \varepsilon{}_2^{-1}
\gamma_{n_{2}} j_2$.

{\textquotedblleft Mixed\textquotedblright} steady state 4, wherein
both lasers are active, exists in the positive quadrant only if
$\varkappa_1 < j_2/j_1$ and $\varkappa_2 < j_1/j_2$. Note, that if
this steady state is physically feasible, then both
{\textquotedblleft pure\textquotedblright} steady states, 1 and 2,
are unstable. The necessary and sufficient conditions for all the
eigenvalues of the Jacobian matrix, evaluated at steady state 4, to
have negative real parts, are, from the Routh--Hurwitz criterion,
\begin{eqnarray}
c_1 & > & 0, \label{hurwitz1} \\
c_1 c_2 - c_3 & > & 0, \label{hurwitz2} \\
(c_1 c_2 - c_3) c_3 - c{}_1^2 c_4 & > & 0, \label{hurwitz3} \\
c_4 & > & 0, \label{hurwitz4}
\end{eqnarray}
where $c_1$, $c_2$, $c_3$ and $c_4$ are the coefficients of the
characteristic polynomial $\lambda^4 +c_1 \lambda^3 +c_2 \lambda^2
+c_3 \lambda +c_4$:
\begin{eqnarray}
c_1 & = & \gamma_{n_1} (p{}_1^{(4)}+1) +\gamma_{n_2} (p{}_2^{(4)}+1), \nonumber\\
c_2 & = & \varepsilon{}_1^{-1} \varepsilon{}_2^{-1} ( \varepsilon_2
\gamma{}_{n_1}^2 p{}_1^{(4)} + \varepsilon_1 \gamma{}_{n_2}^2
p{}_2^{(4)} + ( \varepsilon_2 \varkappa_2 \gamma{}_{n_1}^2 +
\varepsilon_1 \varkappa_1 \gamma{}_{n_2}^2 ) p{}_1^{(4)} p{}_2^{(4)} \nonumber\\
& & + \varepsilon_1 \varepsilon_2 \gamma_{n_1} \gamma_{n_2} (
p{}_1^{(4)} + p{}_2^{(4)} + 1 ) + \gamma_{n_1} \gamma_{n_2} (
\varepsilon_1 \varepsilon_2 - \varkappa_1 \varkappa_2 )
p{}_1^{(4)} p{}_2^{(4)} ), \nonumber\\
c_3 & = & \varepsilon{}_1^{-1} \varepsilon{}_2^{-1} \gamma_{n_1}
\gamma_{n_2} ((\varepsilon_1 \gamma_{n_2} (1 + \varkappa_1
p{}_1^{(4)} + \varkappa_1) + \varepsilon_2 \gamma_{n_1} (1 +
\varkappa_2 p{}_2^{(4)} + \varkappa_2) \nonumber\\
& & - \varkappa_1 \varkappa_2 (\gamma_{n_1} p{}_1^{(4)} +
\gamma_{n_2} p{}_2^{(4)}+ \gamma_{n_1} + \gamma_{n_2}))
p{}_1^{(4)} p{}_2^{(4)} \nonumber\\
& & + \varepsilon_2 \gamma_{n_1} p{}_1^{(4)}
+ \varepsilon_1 \gamma_{n_2} p{}_2^{(4)}), \nonumber\\
c_4 & = & \varepsilon{}_1^{-1} \varepsilon{}_2^{-1} \gamma{}_{n_1}^2
\gamma{}_{n_2}^2 p{}_1^{(4)} p{}_2^{(4)} \left( 1 + \varkappa_1
p{}_1^{(4)} + \varkappa_2 p{}_2^{(4)} - \varkappa_1 \varkappa_2
(p{}_1^{(4)} + p{}_2^{(4)} + 1 ) \right). \nonumber
\end{eqnarray}

Conditions (\ref{hurwitz1}) and (\ref{hurwitz2}) are always
fulfilled. To analyze (\ref{hurwitz3}) and (\ref{hurwitz4}), we
place $\varepsilon_2 = \alpha \varepsilon_2$, $\gamma_{n_2} = \beta
\gamma_{n_2}$, $j_2 = \zeta j_1$ and $\varkappa_2 = \eta
\varkappa_1$, so that the auxiliary parameters $\alpha$, $\beta$,
$\zeta$ and $\eta$ are within $O(1)$. In terms of this assumption,
(\ref{hurwitz3}) can be boiled down, for $\varkappa_1\ll1$, to
\begin{displaymath}
\varepsilon{}_1^{-3} \gamma{}_{n_1}^6 (m_1 \varepsilon_1 - m_2
\varkappa{}_1^2)>0,
\end{displaymath}
where $m_1$ and $m_2$ are positive coefficients depending only on
$j_1$, such that $m_1=O(j{}_1^4)$ and $m_2=O(j{}_1^5)$. It can be
seen that the above inequality and consequently (\ref{hurwitz3})
hold if
\begin{equation}
\label{kappa lt epsilon} \varkappa_1=o(\varepsilon{}_1^{1/2}).
\end{equation}
As is known (\ref{hurwitz3}) guarantees a simple complex conjugate
pair of eigenvalues corresponding to a linearization about steady
state 4 to have negative real part.

Condition (\ref{hurwitz4}) can be shown to yield $\varkappa_1=o(1)$,
however this constraint is weaker than (\ref{kappa lt epsilon}).
Thus (\ref{kappa lt epsilon}) is the stability condition for steady
state 4.

Having regard to a fairly small value of $\varepsilon_1$,
(\ref{kappa lt epsilon}) may be thought to be broken under most
physically meaningful conditions when coupling is not
infinitesimally weak. Hence normally, condition (\ref{hurwitz3}) of
the Routh--Hurwitz criterion is never fulfilled and
{\textquotedblleft mixed\textquotedblright} steady state 4, if any,
is always unstable by growing oscillations.

\begin{figure}
\begin{center}
\subfigure[]{\includegraphics[width=0.48\textwidth]{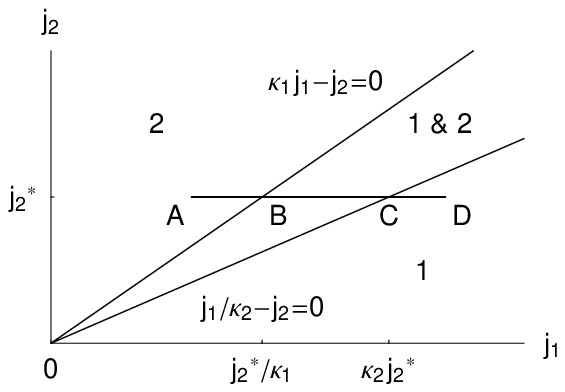}}\hfill
\subfigure[]{\includegraphics[width=0.48\textwidth]{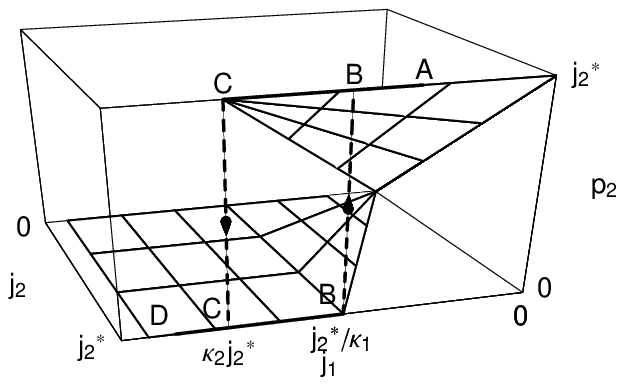}}
\caption{(a) Parameter domain for the number of positive stable
steady states for the model (\ref{coupled lasers}). Only the
positive steady state 1 is stable below the line $ j_1/\varkappa_2 -
j_2 = 0$. Similarly, the only stable steady state is 2 with its
domain of existence above the line $\varkappa_1 j_1 - j_2 = 0$. Both
states coexist in the region of bistability confined by the two
straight lines, realization of either state being a matter of
path-dependency. (b) A cusp catastrophe for the equilibrium states
in the $(p_2,\,j_2,\,j_1)$ parameter space. As $j_1$ increases from
$A$, the path is $ABCCD$ while as $j_1$ decreases from $D$, the path
is $DCBBA$. The projection of the two surfaces onto the
$(j_2,\,j_1)$ plane is given in (a). Two stable equilibria exist
where the overlap is.} \label{hyst}
\end{center}
\end{figure}

At not-too-weak coupling strengths, such that
\begin{equation}
\label{hysteresis condition} \varkappa_1 \varkappa_1 > 1,
\end{equation}
the system being studied is able to exhibit a \emph{hysteresis
effect\/}. (\ref{hysteresis condition}) is obtained combining
(\ref{ss1 stability}) and (\ref{ss2 stability}). Suppose, for
definiteness, that we have a fixed $j_2=j{}_2^*$, and $j_1$
increases from some value less than $j{}_2^*/\varkappa_1$ along the
path $ABCD$ in the $j_2,\,j_1$ parameter space (Fig. \ref{hyst}a).
Then, referring also to (\ref{coupled sol})--(\ref{ss2 stability}),
we see that steady state 2 initially takes place at $A$ with laser 2
on and laser 1 off. This state remains unchanged with $j_1$ until
$C$ in Fig. \ref{hyst}a is reached. For a larger $j_1$ steady state
2 gives up its stability and the system jumps to steady state 1.
Laser 2 becomes dim, while laser 1 takes over. If we now reduce
$j_1$, the system is in steady state 1 and it remains there until
$j_1$ reaches the lower critical value, where there is again only
one stable steady state, at which there is a jump from steady state
1 to steady state 2. In other words as $j_1$ increases along $ABCD$
there is a discontinuous switch from laser 2 to laser 1 at $C$ while
as $j_1$ decreases from $D$ to $A$ there is a discontinuous switch
from laser 2 to laser 1 at $B$. The hysteresis is made possible
thanks to the concurrent stability of both {\textquotedblleft
pure\textquotedblright} equilibria on the interval from $B$ to $C$.
In terms of electronics, such a situation would describe a
\emph{flip-flop circuit\/} having two stable conditions, each
corresponding to one of two alternative input signals. If
(\ref{hysteresis condition}) is not met, then there are no stable
steady states within $BC$. The hysteresis is an example of a cusp
catastrophe which is illustrated schematically in Fig. \ref{hyst}b
where the letters $A$, $B$, $C$ and $D$ correspond to those in Fig.
\ref{hyst}a. Note that Fig. \ref{hyst}a is the projection of the
surface onto the $(j_2,\,j_1)$ plane with the wedge-shaped region
corresponding to the overlap.

It is worth noting that the presence of the second-order loss in
already mentioned model \cite{Vicente2004} leads to basically
similar types of fixed points: two {\textquotedblleft
pure\textquotedblright} and two {\textquotedblleft
mixed\textquotedblright}. However the fundamentally different
coupling scheme induces a quite another bifurcational behavior of
those steady states in regard to bias currents and coupling
strengths.

\section{Two loss-coupled lasers: oscillatory dynamics}
\label{dynamics} When conditions (\ref{ss1 stability}) and (\ref{ss2
stability}) are not met, but instead,
\begin{equation} \label{oscill conditions}
\left\{
\begin{array}{rcl} \varkappa_1 & < & j_2/j_1, \\
\varkappa_2 & < & j_1/j_2,
\end{array}
\right.
\end{equation}
model (\ref{coupled lasers}) has three positive steady states, 2, 3
and 4, none of them being stable. {\textquotedblleft
Mixed\textquotedblright} steady state 4 is unstable by growing
oscillations. In such a case the model would thus be expected to
have a limit cycle in its four-di\-men\-si\-o\-nal phase space
corresponding to sustained oscillations. At sufficiently weak
coupling strengths of order $\varepsilon^{1/2}$, i.\,e. not too far
away from the Hopf bifurcation, where condition (\ref{kappa lt
epsilon}) breaks down, this limit cycle is small and represents a
low-amplitude quasi-harmonic periodic solution. As a practical
matter, the range of such an extremely weak coupling is of less
concern to us than is the range of far more feasible relatively
strong coupling corresponding to well-developed substantially
nonlinear oscillations. We are going to demonstrate that at
not-too-weak coupling strengths and as conditions (\ref{oscill
conditions})take place, system (\ref{coupled lasers}) exhibits
relaxation oscillatory behavior with the two coupled lasers being
antiphase locked.

Since time constants for photons and carriers considerably differ,
four-di\-men\-si\-o\-nal system (\ref{coupled lasers}) is singularly
perturbed. Relatively fast variables are $p_1$ and $p_2$, and slow
variables are $n_1$ and $n_2$. The standard practice of reducing
such systems is adiabatical elimination of the fast variables, when
the left-hand side in the fast equation is replaced by zero, thus
turning this differential equation into an algebraic equation. It is
assumed, that the fast variables quickly relax to their momentary
equilibrium values obtained from the algebraic equations, in which
the slow variables are treated as parameters. {\textquotedblleft
Frozen\textquotedblright} slow variables do not move substantially
in this short adaptation time of the fast variables. The momentary
equilibrium value of the fast variables can thereupon be expressed
by value of the slow variable. The fast variables hastily adapt to
the motion of the slow variables (\emph{order parameters\/}). The
former are entrained by the latter. Chemical physicists who were
first to introduce this technique often refer to it as the
\emph{quasi-steady-state approximation\/} (\emph{QSSA\/}). The
utility of the procedure is that it allows us to reduce the
dimension of the system by retaining only order parameters in the
model. One has to establish the validity of the adiabatical
elimination in each specific case using the recommendations of the
singular perturbation theory. In particular, Tikhonov theorem
\cite{Tikhonov1952} requires quasi-steady state of the fast
equations to be stable.

To replace the derivatives in the first two equations of
(\ref{coupled lasers}),
\begin{equation} \label{adjoint system}
\left\{
\begin{array}{rcl} \dot{p}_1 & = & \varepsilon{}_1^{-1}
\gamma_{n_1} (n_1 - \delta_1
p_1 - \varkappa_2 p_2) p_1, \\
\dot{p}_2 & = & \varepsilon{}_2^{-1} \gamma_{n_2} (n_2 - \delta_2
p_2 - \varkappa_1 p_1) p_2,
\end{array}
\right.
\end{equation}
by zeros and reduce the respective equations to the algebraic system
\begin{equation} \label{algebraic system}
\left\{ \begin{array}{rcl} (n_1-\delta_1 p_1-\varkappa_2 p_2)p_1 & = & 0, \\
(n_2-\delta_2 p_2-\varkappa_1 p_1)p_2 & = & 0,
\end{array} \right.
\end{equation}
in which $n_1$ and $n_2$ are treated as parameters, one has to
ensure stability of quasi-steady states of the fast subsystem
(\ref{adjoint system}).

We anticipate the dynamics of singularly perturbed system
(\ref{coupled lasers}) in the phase space $(p_1,\, p_2,\, n_1,\,
n_2)$ to be consisted of two typical motions: quickly approaching
the {\textquotedblleft slow\textquotedblright} manifold
(\ref{algebraic system}) and slowly sliding along it until a leave
point (where the solution disappears) is reached. After that the
representing point possibly may jump to another local solution of
(\ref{algebraic system}).

Thus, we have to find all quasi-steady states of (\ref{adjoint
system}), distinguish the domains of their stability in the phase
plane $(n_2,\, n_1)$ of the slow subsystem
\begin{equation} \label{degenerate system}
\left\{
\begin{array}{rcl} \dot{n}_1 & = & \gamma_{n_1} \left( j_1 -
(n_1 + 1) p_1 - n_1
\right), \\
\dot{n}_2 & = & \gamma_{n_2} \left( j_2 - (n_2 + 1) p_2 - n_2
\right),
\end{array}
\right.
\end{equation}
and then investigate the dynamics of the complete system
(\ref{coupled lasers}) with piecewise continuous functions.

Subsystem (\ref{adjoint system}) has four quasi-steady states, two
{\textquotedblleft pure\textquotedblright} and two
{\textquotedblleft mixed\textquotedblright} (the slow variables are
deemed to be {\textquotedblleft frozen\textquotedblright}):
\begin{equation}
\label{quasi-steady states}
\begin{array}{rcl}
p{}_{1}^{(\mathrm{qs}1)} & = & n_1/\delta_1,
\qquad p{}_{2}^{(\mathrm{qs}1)} = 0; \\
p{}_{1}^{(\mathrm{qs}2)} & = & 0,
\qquad  p{}_{2}^{(\mathrm{qs}2)} = n_2/\delta_2; \\
p{}_{1}^{(\mathrm{qs}3)} & \approx & ( \varkappa_2n_2 -
\delta_2n_1)/
(\varkappa_1\varkappa_2), \\
p{}_{2}^{(\mathrm{qs}3)} & \approx & ( \varkappa_1n_1 - \delta_1n_2
)/
(\varkappa_1\varkappa_2); \\
p{}_{1}^{(\mathrm{qs}4)} & = & 0, \qquad p{}_{2}^{(\mathrm{qs}4)} =
0.
\end{array}
\end{equation}
(Quasi-steady-state solution 3 is written in line with the
assumption (\ref{kappa gt gamma}).)

\begin{figure}
\begin{center}
\subfigure[]{\includegraphics[width=0.48\textwidth]{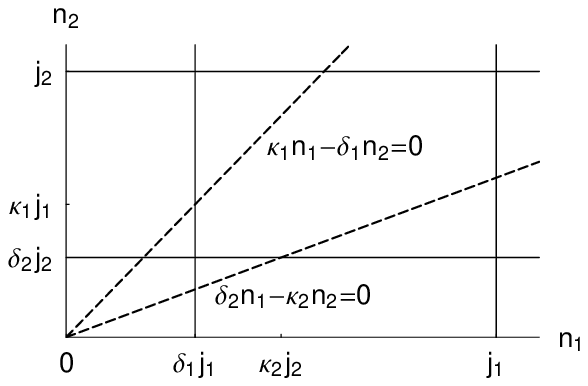}}\hfill
\subfigure[]{\includegraphics[width=0.48\textwidth]{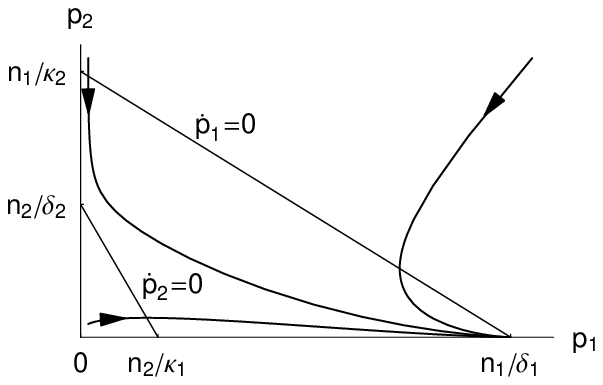}}\\
\subfigure[]{\includegraphics[width=0.48\textwidth]{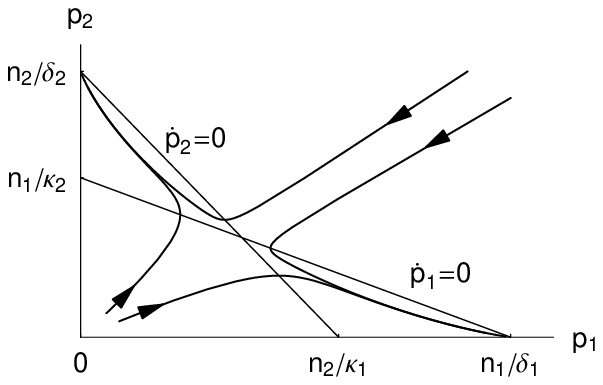}}\hfill
\subfigure[]{\includegraphics[width=0.48\textwidth]{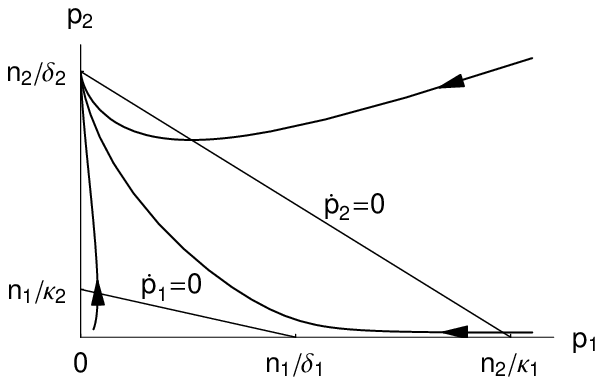}}
\caption{(a) Phase plane of the slow variables (order parameters)
$n_1$ and $n_2$ sectored (by dashed lines) into stability domains of
the corresponding quasi-steady states of the fast subsystem
(\ref{adjoint system}). Both {\textquotedblleft
pure\textquotedblright} quasi-steady states, each corresponding to
the situation when either of the two lasers is on, coexist within
the opening of the angle formed by dashed lines (\ref{stability
qss1}) and (\ref{stability qss2}). Lines $n_1 = \delta_1 j_1 $ and
$n_2 = j_2$ are respective nullclines $\dot{n}_1 = 0$ and $\dot{n}_2
= 0$ of the piecewise system (\ref{n1 piecewise1})--(\ref{n2
piecewise1}). Lines $n_1 = j_1$ and $n_2 = \delta_2 j_2$ mean the
same for the system (\ref{n1 piecewise2})--(\ref{n2 piecewise2}).
Intersections of the nullclines are equilibria of the associated
piecewise slow subsystems, and they must be outside the
abovementioned opening to allow for the relaxation oscillations. (b)
Phase portrait of (\ref{adjoint system}) for the case in which
{\textquotedblleft pure\textquotedblright} quasi-steady state 1 is
the only stable solution. (c) Phase portrait of (\ref{adjoint
system}) for the flip-flop case in which both {\textquotedblleft
pure\textquotedblright} quasi-steady states are stable nodes and
coexist being separated by a saddle point. Realization of either
state depends on the initial conditions. (d) Phase portrait of
(\ref{adjoint system}) for the case in which {\textquotedblleft
pure\textquotedblright} quasi-steady state 2 is the only stable
solution.} \label{qss}
\end{center}
\end{figure}

{\textquotedblleft Pure\textquotedblright} quasi-steady state 1 is a
stable node everywhere below the line
\begin{equation}
\label{stability qss1} \varkappa_1n_1-\delta_1n_2=0
\end{equation}
in the parametric plane $n_2$, $n_1$ of the slow variables (Fig.
\ref{qss}a). It is realizable in two different phase portraits shown
in Figs. \ref{qss}b and \ref{qss}c.
By the same token {\textquotedblleft pure\textquotedblright}
quasi-steady state 2 is a stable node everywhere above the line
\begin{equation}
\label{stability qss2} \varkappa_2n_2 - \delta_2 n_1 = 0
\end{equation}
(Fig. \ref{qss}a), and it is featured by the phase portraits in
Figs. \ref{qss}c and \ref{qss}d.

Both {\textquotedblleft pure\textquotedblright} quasi-steady states
can coexist within the opening of the angle formed by lines
(\ref{stability qss1}) and (\ref{stability qss2}) in Fig.
\ref{qss}a:
\begin{equation}
\label{opening} \delta_2 n_1 /\varkappa_2 < n_2 < \varkappa_1
n_1/\delta_1.
\end{equation}
The opening shrinks as coupling strengths get weaker. In this
flip-flop domain the two {\textquotedblleft pure\textquotedblright}
quasi-steady states, both stable, are being separated by
{\textquotedblleft mixed\textquotedblright} quasi-steady state 3 of
a saddle type (Fig. \ref{qss}a).

As to {\textquotedblleft mixed\textquotedblright} (trivial)
quasi-steady state 4, it is always an unstable node.

Let us assume that {\textquotedblleft pure\textquotedblright}
quasi-steady state 1 is initially stable, the population inversions
$n_1$ and $n_2$ are somewhere within the domain $\varkappa_1 n_1 -
\delta_1 n_2 > 0$, and also $n_1(0) \gg \delta_1 j_1$. While $n_1$
remains much greater than $\delta_1j_1$, the dynamics of the slow
variables (treated as bifurcation parameters in reference to the
fast variables) is governed by a system of two independent equations
\begin{eqnarray}
\dot{n}_1 & = & \gamma_{n_1} \left( j_1 - n_1 (n_1 + 1 )/
\delta_1 - n_1 \right), \label{n1 piecewise1} \\
\dot{n}_2 & = & \gamma_{n_2} \left( j_2 - n_2 \right). \label{n2
piecewise1}
\end{eqnarray}
This system has (stable) steady state
\begin{equation}
\label{n steady state 1} n{}_1^{(1)} = \delta_1 j_1 +
O(\delta{}_1^2), \qquad n{}_2^{(1)} = j_2,
\end{equation}
and the representing point will tend to reach it. System (\ref{n1
piecewise1})--(\ref{n2 piecewise1}) also allows for distinguishing
fast and slow variables. Due to small value of $\delta_1$, equation
(\ref{n1 piecewise1}) is roughly $\delta{}_1^{-1}$ faster than
equation (\ref{n2 piecewise1}). Therefore $n_1$ relatively quickly
relaxes to $n{}_1^{(1)}$, $n_2$ being practically {\textquotedblleft
frozen\textquotedblright}. In other words, the representing point
first arrives at the {\textquotedblleft slow\textquotedblright}
nullcline $\dot{n}_1 = 0$, given by $n_1 \approx \delta_1 j_1$. If
$\delta_1$ were large enough to damp the intrinsic oscillations of
laser 1, the representing point would further slowly slide along
that nullcline, calmly tending to $j_2$. Still small value of
$\delta_1$ complicates the picture, and in the immediate vicinity of
$n{}_1^{(1)}$ the system gets trapped into a stable focus with
respect to variables $p_1$ and $n_1$. Here QSSA ceases to be true,
and one can no longer substitute $p_1$ by its quasi-steady-state
value $p{}_1^{(\mathrm{qss}1)}$ . Instead of (\ref{n1 piecewise1})
we have to write down a pair of equations
\begin{equation} \label{1st laser}
\left\{
\begin{array}{rcl} \dot{p}_1 & = & \varepsilon{}_1^{-1} \gamma_{n_1}
\left( n_1 - \delta_1 p_1 \right) p_1, \\
\dot{n}_1 & = & \gamma_{n_1} \left( j_1 - (n_1 + 1) p_1 - n_1
\right).
\end{array}
\right.
\end{equation}

\begin{figure}
\begin{center}
\includegraphics{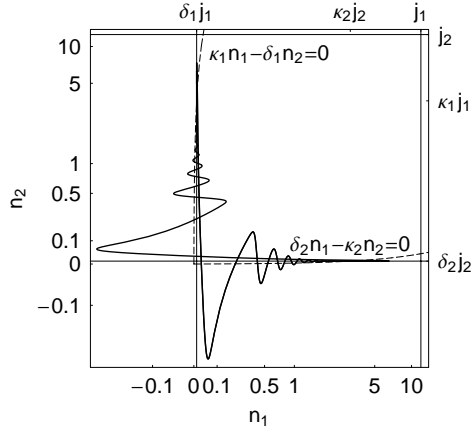}
\caption{Logarithmically stretched-out neighborhood of point
$(\delta_1 j_1,\,\delta_2 j_2)$ in the phase plane of the order
parameters (cf. Fig. \ref{qss}a) and the projection of the limit
cycle. In this numerical example parameters of the two lasers are
identical with the exception of the pumps chosen to be $j_1 =
11\ldotp 9$ and $j_2 = 12\ldotp 5$. Coupling strengths are
$\varkappa_1 = 0\ldotp 30$ and $\varkappa_2 = 0\ldotp 25$. Note
transverse damped vibrations of the representing point whenever it
moves along either of two nullclines $\dot{n}_1 = 0$ and $\dot{n}_2
= 0$.} \label{stretched-out}
\end{center}
\end{figure}

System (\ref{1st laser}) is identical to the rate equations
(\ref{basic eqs}) of an uncoupled laser and in essence describes
underdamped intrinsic oscillations. In the plane of the slow
variables these oscillations manifest themselves in damped
transverse fluctuations superimposed on the independent vertical
motion along the nullcline $\dot{n}_1 = 0$ toward $n_2^{(1)} = j_2$
(Fig. \ref{stretched-out}). Thus, $n_2$ is the actual order
parameter in this area of the complete four-di\-men\-si\-o\-nal
phase space. If steady state 1 (given by (\ref{n steady state 1}))
lies beyond the area $\varkappa_1 n_1 - \delta_1 n_2 > 0$ (Fig.
\ref{qss}a), or the same, if the first inequality of (\ref{oscill
conditions}) holds true, then the representing point would
inevitably touch the upper boundary of the domain (\ref{opening}) at
a point $(\delta_1 j_1,\,\varkappa_1 j_1)$ before approaching the
vicinity of steady state 1. On the boundary given by equation
(\ref{stability qss1}), {\textquotedblleft pure\textquotedblright}
quasi-steady state 1 for the entrained variables merges with
{\textquotedblleft mixed\textquotedblright} saddle quasi-steady
state 3 and loses its stability. Laser 1 instantly switches off, and
the alternative {\textquotedblleft pure\textquotedblright}
quasi-steady state 2 becomes stable, with laser 2 being on.

In terms of four-di\-men\-si\-o\-nal phase space of the complete
system (\ref{coupled lasers}), the representing point is now in the
other stable hyperplane of the {\textquotedblleft
slow\textquotedblright} manifold (\ref{algebraic system}). Slow
subsystem sliding along this alternative branch obeys the equations
\begin{eqnarray}
\dot{n}_1 & = & \gamma_{n_1} \left( j_1 - n_1 \right),
\label{n1 piecewise2} \\
\dot{n}_2 & = & \gamma_{n_2} \left( j_2 - n_2 (n_2 + 1)/\delta_2 -
n_2 \right), \label{n2 piecewise2}
\end{eqnarray}
with the initial conditions $n_1(0) = n_1^{(1)} = \delta_1 j_1$ and
$n_2(0) = \varkappa_1 j_1$. System (\ref{n1 piecewise2})--(\ref{n2
piecewise2}) has steady state
\begin{equation}
\label{n steady state 2} n_1^{(2)} = j_1, \qquad n_2^{(2)} =
\delta_2 j_2 + O(\delta{}_2^2),
\end{equation}
that is stable. Variable $n_2$ quickly approaches the nullcline
$\dot{n}_2 = 0$ given by $n_2 \approx \delta_2 j_2$, and then starts
to oscillate about it according to the equations
\begin{equation}
\label{2nd laser} \left\{
\begin{array}{rcl} \dot{p}_2 &
= & \varepsilon{}_2^{-1} \gamma_{n_2} \left( n_2 - \delta_2 p_2 \right) p_2, \\
\dot{n}_2 & = & \gamma_{n_2} \left( j_2 - (n_2 + 1) p_2 - n_2
\right).
\end{array}
\right.
\end{equation}
Equations (\ref{2nd laser}) describe underdamped intrinsic
oscillations of laser 2. At the same time $n_1$ (which is now the
order parameter) relatively slowly tends to $j_1$ along the
nullcline $\dot{n}_2 = 0$. Again, if steady state (\ref{n steady
state 2}) is located below line (\ref{stability qss2}), or equally,
if the second inequality of (\ref{oscill conditions}) holds true,
then the representing point would certainly touch boundary
(\ref{stability qss2}) at a point $(\delta_2 j_2,\,\varkappa_2
j_2)$, whereupon {\textquotedblleft pure\textquotedblright}
quasi-steady state 2 for the fast variables would merge with saddle
quasi-steady state 3 without fail and lose its stability in favor of
{\textquotedblleft pure\textquotedblright} quasi-steady state 1. The
system returns to the first branch, and the oscillatory cycle gets
closed.

\section{Discussion}
\label{discussion} Thus, under conditions (\ref{oscill conditions})
system (\ref{coupled lasers}) features sustained relaxation
oscillations (Fig. \ref{time profiles}). The coupled lasers turn out
to be antiphase locked. In its principle of operation, the
considered system resembles \emph{astable multivibrator}, known to
be an electronic circuit that oscillates between its two states,
neither of which is stable, generating a continuous flow of square
edge pulses.

\begin{figure}
\begin{center}
\subfigure[]{\includegraphics[width=0.48\textwidth]{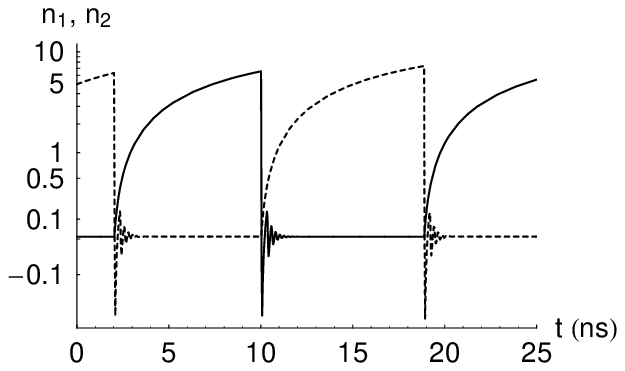}}\hfill
\subfigure[]{\includegraphics[width=0.48\textwidth]{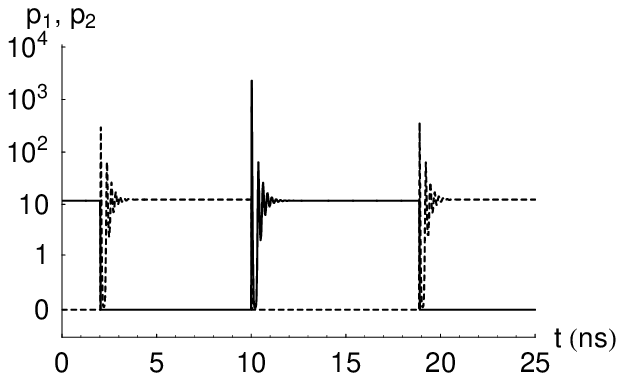}}
\caption{Time profiles of oscillating population inversions (a) and
photon numbers (b) in two loss-coupled lasers. Solid: laser 1,
dashed: laser 2. Numerical values of the parameters are those
mentioned in the caption to Fig. \ref{stretched-out}.} \label{time
profiles}
\end{center}
\end{figure}

It should be mentioned that quite similar antiphase synchronization
patterns have been previously found (both numerically and
experimentally) by T.~Baer \cite{Baer1986} and K.~Wiesenfeld,
C.~Bracikowski, G.~James and R.~Roy \cite{Wiesenfeld1990} in a
different system featured by two coupled longitudinal modes
oscillating in Nd:YAG laser with an intracavity-doubling crystal.

The population inversions, $n_1$ and $n_2$, demonstrate saw-tooth
periodical pulses. Oscillation range for the population inversions
remains finite, and, what is important, does not depend on
$\delta_1$ and $\delta_2$. The respective amplitudes for $n_1$ and
$n_2$ are of orders of $\varkappa_2 j_2$ and $\varkappa_1 j_1$.

Photon numbers, $p_1$ and $p_2$, inside the cavities change
periodically between quiescence and short giant spikes. The
magnitude of power output spikes, in contrast to carrier's pulses,
tends to infinity as $\delta_{1,\,2} \rightarrow 0$, in view of
equations (\ref{quasi-steady states}).

\begin{figure}
\begin{center}
\subfigure[]{\includegraphics[width=0.48\textwidth]{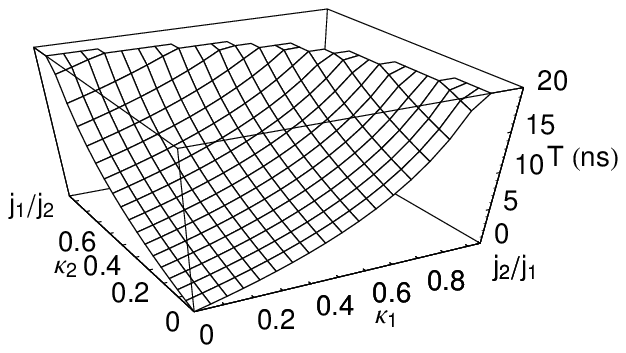}}\hfill
\subfigure[]{\includegraphics[width=0.48\textwidth]{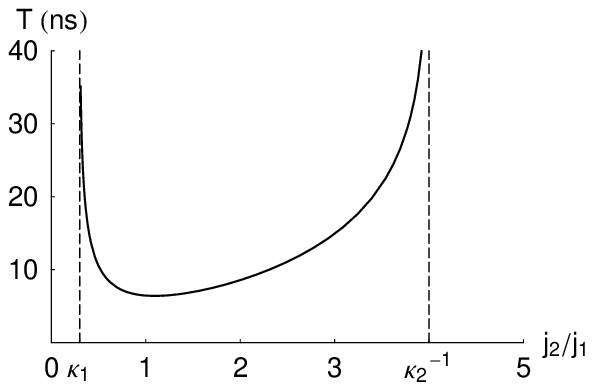}}
\caption{The period of synchronous oscillations versus coupling
strengths (a) and pumps ratio (b). Numerical values of the
parameters are those mentioned in the caption to Fig.
\ref{stretched-out}.} \label{period graphs}
\end{center}
\end{figure}

Times of motion over the either branch of the {\textquotedblleft
slow\textquotedblright} manifold (\ref{algebraic system}), $\tau_1$
and $\tau_2$, add up to give a period of oscillations, $T$. Those
times are predominantly determined by dynamics of the order
parameters $n_1$ and $n_2$, and, to a zeroth approximation in
$\varepsilon_1$ and $\varepsilon_2$, can be found as solutions of
the equations of motion (\ref{n1 piecewise2}) and (\ref{n2
piecewise1}) with respective boundary conditions $n_1(0) = 0$, $n_1
(\tau_1) = \varkappa_2 j_2$, and $n_2(0) = 0$, $n_2 (\tau_2)=
\varkappa_1 j_1$. Therefore we obtain a quite simple estimate for
the period:
\begin{eqnarray}
T & = & \tau_1 + \tau_2 = \gamma{}_{n_1}^{-1} \displaystyle
\int\nolimits_{0}^{\varkappa_2 j_2}\frac{d\xi}{j_1 - \xi} +
\gamma{}_{n_2}^{-1} \displaystyle
\int\nolimits_{0}^{\varkappa_1 j_1}\frac{d\xi}{j_2 - \xi} \nonumber \\
& = & \gamma{}_{n_1}^{-1} \ln\frac{1}{1 - \varkappa_2 j_2/j_1} +
\gamma{}_{n_2}^{-1} \ln\frac{1}{1 - \varkappa_1 j_1/j_2}\,.
\label{relax period}
\end{eqnarray}
The typical temporal scale of the oscillations is determined by
carrier population inversion time constants and turns to be measured
in nanoseconds (which is much longer than underdamped oscillations
of a single laser). It is interesting, that according to (\ref{relax
period}) the period depends on the ratio of the pump currents,
$j_2/j_1$, rather than on each of the two currents individually, and
completely does not depend on concrete values of $\delta_1$ and
$\delta_2$. At weak coupling the period linearly shortens, but tends
to infinity whenever either $\varkappa_1$ or $\varkappa_2^{-1}$
approaches $j_2/j_1$ (Fig. \ref{period graphs}a). Given
$\varkappa_1$ and $\varkappa_2$, such that $\varkappa_1\varkappa_2 <
1$, the ratio $j_2/j_1$ has to be confined between $\varkappa_1$ and
$\varkappa_2^{-1}$ (Fig. \ref{period graphs}b). Within that area the
period relatively weakly depends on the ratio of the two pumps.

In all likelihood, the limit cycle would persist in our model of
coupled lasers even if the second-order cavity loss is not merely
small but absent whatsoever. However in such a case the quasi-steady
state approximation technique is inapplicable since at
$\delta_{1,\,2} = 0$ fast subsystem (\ref{adjoint system}) has no
finite {\textquotedblleft pure\textquotedblright} equilibria.
Generally speaking, as $\delta_{1,\,2}$ becomes of order
$\varepsilon_{1,\,2}$ equations (\ref{n1 piecewise1}) and (\ref{n2
piecewise2}) can no longer be considered slow and the approximate
estimate (\ref{relax period}) loses its accuracy. QSSA yields its
best accuracy for $\varepsilon_{1,\,2} \ll \delta_{1,\,2} \ll 0$.

Earlier, B.A.~Nguyen and P.~Mandel \cite{Nguyen1999} have shown (in
the framework of their model) that equally pumped lasers become
unstable for much smaller values of the loss cross-coupling than
unequally pumped lasers. Their result has to do with the onset of
oscillations. Our model deals with well developed high-amplitude
nonlinear oscillations and therefore allows to predicts somewhat
similar related to quenching: synchronous oscillations in equally
pumped lasers are being quenched for smaller values of the greater
of two coupling strengths than in unequally pumped lasers. Indeed,
it follows from conditions (\ref{oscill conditions}) that if pumps
are equal, $j_1 = j_2$, then both coupling strengths, $\varkappa_1$
and $\varkappa_2$, must be less than unity for the synchronous
oscillations to go on. In case of the different pumps, however, one
of the coupling strengths may be greater than unity.

The most intriguing feature of the considered model is that each of
the two lasers by itself does not lase, however in interaction, when
coupled in a nonlinear way, the resulting system is shown to have
sustained oscillations. As far back as in early 1970s, S.~Smale
\cite{Smale1974} constructed an abstract mathematical example of a
cell modeled by the chemical kinetics of four metabolites, $x_1,
\cdots, x_4$, such that the reaction equations $\dot {\mathbf x} =
\mathbf R(\mathbf x)$ for the set of metabolites, $\mathbf x = (x_1,
\cdots, x_4)$, had a globally stable equilibrium. The cell is
{\textquotedblleft dead\textquotedblright} in that the
concentrations of its metabolites always tend to the same fixed
levels. When two such cells are coupled by linear diffusion terms of
the form $\mathbf M (\mathbf x_2 - \mathbf x_1)$, where $\mathbf M$
is a diagonal matrix with the elements $\mu_k \delta_{kl}$, however,
the resulting equations are shown to have a globally stable limit
cycle. The concentrations of the metabolites begin to oscillate, and
the system becomes {\textquotedblleft alive\textquotedblright}. In
Smale's words:
\begin{quote}
There is a paradoxical aspect to the example. One has two dead
(mathematically dead) cells interacting by a diffusion process which
has a tendency in itself to equalize the concentrations. Yet in
interaction, a state continues to pulse indefinitely.
\end{quote}
Smale also remarks that \textquoteleft it is more difficult to
reduce the number of chemicals to two or even three.\textquoteright

Equations (\ref{coupled lasers}) may be interpreted in biological
terms if we assume $p_1$, $p_2$ and $n_1$, $n_2$ to be respectively
species (predators) and nutrients (preys) consumed by the species.
The corresponding nutrients are fed into the system with some
constant rates $j_1$ and $j_2$. In such a case the model describes
\emph{interference interspecific competition\/} between two
predators belonging to different species -- competition that does
not act through the utilization of a nutritious resource, but
instead involves direct interaction between the competitors (e.g.
through aggressive behavior). Coupling strengths $\varkappa_1$ and
$\varkappa_2$ just reflect the intensity of interspecific
competition. As distinct from the Smale's example, coupling is
nonlinear and this makes sustained synchronous oscillations possible
for fewer number of variables. Constants $\delta_1$ and $\delta_2$
(sometimes referred to as \emph{Verhulst parameters\/} in ecology)
are responsible for \emph{intraspecific\/} competition resulting in
a reduction of population growth rate as population density
increases. Unfortunately the ecological analogy is limited because
of at least two important reasons.

First, species usually influence other species less strongly than
they do their conspecifics, in other words, interspecific
competition is typically dominated by intraspecific competition:
\begin{equation} \label{intra gt inter}
\varkappa_{1,\,2} < \delta_{1,\,2}.
\end{equation}
In our case of loss-coupled lasers, second-order cavity losses are
almost negligible in comparison with the coupling (recall
(\ref{kappa gt gamma})). As early as in 1930s G.F.~Gause and
A.A.~Witt \cite{Gause1935} considered two competing species
described by a pair of equations like (\ref{adjoint system}) with
constant $n_1$ and $n_2$ (which implies the abundance of food
resources) and showed that when conditions (\ref{intra gt inter})
are met, the two species would coexist. In our model the two photon
populations neither coexist concurrently, nor exclude each other
forever. Nevertheless we can say that the two populations coexist in
different temporal niches, in the manner of time sharing.

Second, time scales are usually inverted in ecosystems as opposed to
laser. That is to say, food is consumed by species rapidly, i.e.
\begin{displaymath}
\gamma_{n_{1,\,2}} \gg \gamma_{p_{1,\,2}}
\end{displaymath}
is the common case in ecology. In our model slowness of carrier
population relative to photon population is essential for the
oscillations to occur because it provides the necessary inertia to
the system.

It is notable in this connection that recently M.-Y.~Kim, R.~Roy,
J.L.~Aron, T.W.~Carr and I.B.~Schwartz \cite{Kim2005} have found a
case of complete analogy between coupled lasers and coupled living
populations. It is shown that the rate equations for two lasers
coupled optoelectronically through their pump currents are identical
to the rate equations for two migration-coupled infective human
populations. From the ecological perspective, the model being
discussed and ours describe different types of competition. Model
\cite{Kim2005} corresponds to \emph{trophic interspecific
competition}, because competitors are allowed to affect food sources
of each other. In that model coupling is brought about through a
slow variable (carriers/susceptible individuals), while in model
(\ref{coupled lasers}) the lasers are coupled through a fast
variable (photons). For this reason the two models exhibit different
behavior. In particular, the presence of time delay in coupling is
prerequisite to quasi-harmonic self-pulsing regime in model
\cite{Kim2005}. Synchronous anti-phase oscillations in our model do
not require such a delay, although introducing it to the equations
has much potential for further studying the system.

The author expresses his gratitude to R.~Roy and I.B.~Schwartz for
fruitful discussion.




\begin{thebibliography}{00}
\bibitem{relax} An unhappy term {\textquotedblleft relaxation oscillations\textquotedblright} is quite often used in
reference to those quasi-harmonic oscillations. For the avoidance of
confusion, it seems reasonable to reserve this term to highly
nonlinear discontinuous (jerky)
oscillations.
\end{thebibliography}
\end{document}